\documentclass[twocolumn,showpacs,amsmath,amssymb, nobibnotes, aps, prl,showkeys]{revtex4}
\usepackage{graphicx}
\usepackage{dcolumn}
\usepackage{bm}
\usepackage{docs}
\usepackage[dvipsnames]{xcolor}
\usepackage{newtxtext,newtxmath}
\usepackage[T1]{fontenc}

\begin{document}

\title{Interpreting correlated observations of cosmic rays and gamma-rays from Centaurus A with a proton blazar inspired model}

\author{Prabir Banik$^{1,2}$\thanks{Email address: pbanik74@yahoo.com}, Arunava Bhadra$^{2}$\thanks{Email address: aru\_bhadra@yahoo.com} and Abhijit Bhattacharyya$^{3}$\thanks{Email address: abhattacharyyacu@gmail.com}}
\affiliation{ $^{1}$Department of physics, Surendra Institute of Engineering $\&$ Management, Dhukuria, Siliguri, West Bengal, India 734009}
\affiliation{ $^{2}$High Energy $\&$ Cosmic Ray Research Centre, University of North Bengal, Siliguri, West Bengal, India 734013}
\affiliation{ $^{3}$Department of physics, University of Calcutta, 92 APC Road, Kolkata, West Bengal, India 700009}

\begin{abstract}
The nearest active radio galaxy Centaurus (Cen) A is a gamma-ray emitter in GeV to TeV energy scale. The High Energy Stereoscopic System (H.E.S.S.) and non-simultaneous Fermi-LAT observation indicate an unusual spectral hardening above few GeV energies in the gamma-ray spectrum of Cen A. Very recently the H.E.S.S. observatory resolved the kilo parsec (kpc)-scale jets in Centaurus A at TeV energies. On the other hand, the Pierre Auger Observatory (PAO) detects a few ultra high energy cosmic ray (UHECR) events from Cen-A. The proton blazar inspired model, which considers acceleration of both electrons and hadronic cosmic rays in AGN jet, can explain the observed coincident high energy neutrinos and gamma rays from Ice-cube detected AGN jets. Here we have employed the proton blazar inspired model to explain the observed GeV to TeV gamma-ray spectrum features including the spectrum hardening at GeV energies along with the PAO observation on cosmic rays from Cen-A. Our findings suggest that the model can explain consistently the observed electromagnetic spectrum in combination with the appropriate number of UHECRs from Cen A.
\end{abstract}

\maketitle

\section{Introduction}
The origin of ultrahigh-energy cosmic rays (UHECRs) is one of the long-standing unsolved problems of physics. It is generally believed that the sources of UHECRCs can identify through observations of anisotropy in the arrival direction of UHECRs. The Pierre Auger collaboration reported that arrival directions of the 13 events above 55 EeV energies out of a total of 69 events in the whole sky are consistent with the sky location of a circular window of radius $18^{\circ}$ centered on nearest active galaxy Centaurus (Cen) A, while 3.2 events were expected from the isotropic background \cite{Abreu10}. Moreover, two of the events appeared within less than $3^{\circ}$ about the sky location of the radiogalaxy \cite{Abreu10}. In 2017, the PAO reported an increased correlation of 19 events above 58 EeV with a sky location of a circular window of radius $15^0$ centered on Centaurus A, where 6 events are expected to be isotropic background with a statistical significance of $3.1\sigma$ \cite{Giaccari17}. The observed association of UHECRs with an AGN suggests that AGN may indeed be one of the long-sought sources of very-high-energy cosmic rays. 

The Cen A is classified as one of the radio galaxies of Fanaroff-Riley type I (FR-I) \cite{Fanaroff74} which is thought to correspond to BL Lacertae (BL Lac) objects but the jet is oriented at a large angle to the line of sight. In the active galactic nucleus unification scheme, BL Lac object is a subclass of blazars with relativistic jets aligned along the line of sight \cite{Urry95} with the inner jet is traveling with Lorentz factors of order ten \cite[e.g.][]{Chiaberge01}. The inclination angle of jet of Cen A with the line of sight has been estimated in the range of ($50-80^{\circ}$) \cite{Tingay01} and ($12-45^{\circ}$) \cite{Muller14}. The jet in Cen A becomes subluminal with an apparent speed of $0.5c$ on the hundred parsec-scale distance \cite{Hardcastle03}. Therefore, the bulk motion of jet is expected to be strongly decelerates between the sub-pc and kpc distance scales \cite{Bednarek19}.

Cen A is recognized for interesting radio structure on several size scales, having two giant outer lobes extending over $\sim 10^o$ on the sky and it aligned primarily in the north-south direction \cite{Feain11}. The high-resolution radio very long baseline interferometry (VLBI) observations resolved the innermost small radio core of size $\sim 3\times 10^{16}$ cm and counter-jet features on subparsec scales \cite{Kellermann97,Horiuchi06}. A strong, well-collimated, one-sided, kpc-scale (up to $\sim 4.5$ kpc in projection) jet has been resolved into several bright knots and diffuse emissions by Chandra X-ray observations. This x-ray emission plausibly caused by synchrotron emission \cite{Kraft02,Hardcastle03}. The Cen A was detected by all instruments on board the Compton Gamma-Ray Observatory i.e, Burst and Transient Source Experiment (BATSE), Oriented Scintillation Spectrometer Experiment (OSSE), imaging COMPton TELescope (COMPTEL), and Energetic Gamma Ray Experiment Telescope (EGRET) during the period of 1991$-$1995 in the few tens of keV to GeV energy range of its electromagnetic spectrum. These observations found a second-hump in the electromagnetic (EM) SED of Cen A at an energy of $\sim$ 0.1 MeV \cite{Kinzer95, Steinle98, Sreekumar99}. A single-zone synchrotron self-Compton (SSC) model can successfully interpret the electromagnetic SED ranging from the radio band to the gamma-ray band \cite{Chiaberge01} but the model predicts lower flux than the observed one at very-high-energies (VHE).

Based on 115 hr of observation performed from April 2004 to July 2008, the High Energy Stereoscopic System (H.E.S.S.) discovered Cen A in VHE gamma-rays with a statistical significance of $5 \sigma$ from the region including the radio core and the inner kiloparsec-scale jets \cite{Aharonian09}. The H.E.S.S. collaboration recently reported a firm detection of the Cen A gamma-ray core with a statistical significance of $12\sigma$ based on 213 hours of total exposure time performed from 2004 to 2010 \cite{Abdalla18}. The current spectral investigation of the entire data set reveals a photon index of $\Gamma_{\gamma} = 2.52 \pm 0.13_{stat} \pm 0.20_{sys}$ in the energy range of 250 GeV$-$6 TeV \cite{Abdalla18}. In the H.E.S.S. data set, significant variability was seemed to be absent in recent scenario \cite{Abdalla18}. 

Succeeding survey observations by the Fermi Large Area Telescope (LAT) has reported the detection of high energy (100 MeV$-$100 GeV) gamma-rays from the gamma-ray core (i.e., within $0^{\circ}_{.}1$) \cite{Abdo10a}. The analyses with extended Fermi-LAT data sets revealed clear evidence of substantial spectral break above a few GeV energies in 2013 \cite{Sahakyan13}, and a spectral hardening by $\Delta \Gamma_{\gamma} \approx 0.4\pm 0.1$ at gamma-ray energies above the break energy of $\approx 2.8$ GeV at a level of $\approx 4.0\sigma$ was reported recently \cite{Abdalla18}. This hardening suggests that the high energy gamma-ray emission above the break energy demands an additional gamma-ray emitting spectral component to resolve. On the other hand, observed association of UHECRs with Cen A as detected by PAO strongly suggests that the observed high energy gamma-ray spectrum above the break energy may originate in hadronic interactions in the jet of Cen A if such identified correlation is genuine.

A single-zone SSC model can describe the observed parsec-scale emission spectrum from the core of Cen-A up to $\sim 10$ GeV energies \cite{Abdo10a} but it cannot explain the higher energy (TeV) emission detected by H.E.S.S. during 2004$-$2008. The observations of the VHE component above GeV energies have been interpreted with different production scenarios, namely, the two-zone homogeneous SSC model \cite{Abdalla18}, the photohadronic interaction model ($p\gamma$) \cite{Sahu12,Petropoulou14,Fraija14}, photo-disintegration of accelerated iron nuclei \cite{Joshi13}, a two zone leptonic with photohadronic interaction model \cite{Joshi13} and hadronic ($pp$) interactions in the giants lobes of Cen A \cite{Fraija12} etc. Very recently H.E.S.S. collaboration has been able to resolve the kpc-scale jet of Cen A in TeV energies; they found that instead of a point like emission an elliptically elongated Gaussian  model describes their observation much better \cite{Hess20}. The H.E.S.S. Collaboration has explained their new observation with the external Compton (EC) of accelerated electrons by the different types of soft radiation in a relativistically moving kpc-scale jet \cite{Hess20}. On the other hand \cite{Bednarek19} demonstrated that the TeV gamma-ray emissions from kiloparsec-scale jet can be explained by the  EC of accelerated electrons with inner jet photons. 

In the couple of previous works \cite{Banik19, Banik20}, we have developed a proton blazar inspired model which can describe consistently the observed high energy gamma rays and neutrino signal from all the three Icecube blazars  (i.e, TXS 0506+056, PKS 0502+049, GB6 J1040+0617) \cite{Banik19, Banik20} assuming that the association of the observed neutrino events with the corresponding blazars at the flaring stage are genuine. Basically the model explains the high energy gamma-ray and neutrino observations based on interactions of shock accelerated protons in AGN jet with the cold protons present in the system. Like several other models, the proton blazar inspired model also assumes that protons and nuclei are accelerated along with electrons to relativistic energies in the AGN jets. In this framework, the lower part of high energy component (up to GeV energies) of the observed electromagnetic radiation is explained in terms of inverse Compton (IC) scattering of accelerated electrons with the seed synchrotron photons co-moving with the AGN jet and the observed high energy gamma rays and neutrinos are interpreted through the interaction of the protons/nuclei with cold protons present in the system. 

In the present work, we would like to introduce a two emission region scenarios in the jet under the framework of the proton blazar inspired model to explain the correlated cosmic rays and gamma ray observation. The two emission regions that we have considered in our model are i) the core emission region (subpersec to persec jet scale), where synchrotron self-Compton mechanism dominantly describes the spectral behavior of the observed radio to a few GeV energy EM SED from the central part of Cen A; ii) the kpc-scale jet region, the EC of accelerated electrons with inner jet photons and interaction of accelerated cosmic rays with cold protons mainly lead the spectral behavior of gamma rays in the GeV-TeV energies from the kpc-scale jet of Cen A. The cosmic ray flux and energy spectrum used as an input to reproduce observed gamma rays not only explains the PAO excess but also correctly gives the energy spectrum of the observed cosmic rays. We shall consider a mixed composition for Cen A accelerated cosmic rays. We would also like to estimate the corresponding neutrino fluxes and their detection likelihood by IceCube neutrino observatory.

The arrangement of this paper is as follows: In the next section, we shall explain the methodology for estimating the gamma-ray and neutrino fluxes produced in the interaction of cosmic rays with an ambient matter in the AGN jet under the framework of the adopted proton blazar inspired model. In Sec. 3, we shall evaluate the fluxes of hadronically produced gamma-rays and neutrinos over the GeV to TeV energy range consistently with observed UHECR events from Cen A, and we will compare our computations with the observed spectra. We shall discuss the findings of the present work in Sec. 4 and conclude finally in the same section.

\section{Methodology for gamma-ray and neutrino flux estimation}
In almost all hadronic models of AGN jet, it is usually thought that the relativistic jet material is composed of relativistic protons (p) and electrons (e$^{-}$). But the overall jet composition of AGN is an unresolved issue so far. A hybrid simulation study which demonstrated that only a small fraction of protons of the system (roughly 4\%) are accelerated to non-thermal energies at diffusive shocks, suggests the existence of cold protons in the blazar jets \cite{Caprioli15}. In principle, cold (non-relativistic) protons that arose from charge neutrality condition, are also assumed to exist in the blazar jets to explain high energy gamma-rays and neutrinos as described in the adopted proton blazar inspired model \cite{Banik19, Banik20}. In the present work we explain the observed high energy gamma-rays and UHECR events from the radio galaxy Cen A based on the proton blazar inspired model \cite{Banik19}.

\subsection{The core emission with SSC model}
The EM SED from central part of Cen A exhibits multiple peaks; the lower energy peak at 0.1 eV (infrared), and the higher energy peak at 170 keV (soft gamma-rays) are generally explained by the leptonic models assuming a compact core emission region \cite{Chiaberge01,Abdo10a,Petropoulou14}. Here we consider a moving spherical blob of size $R_b'$ (primed variables for the jet frame) in the AGN jet is responsible for the non-thermal emission and contains a tangled magnetic field of strength $B'$. If $\Gamma_j = 1/\sqrt{1-\beta_j^2}$ be the bulk Lorentz factor of the blob then Doppler factor of the moving blob can be written as $\delta_{ob} = \Gamma_j^{-1}(1-\beta_j\cos\theta)^{-1}$ where $\theta$ is the angle between the line of sight and the jet axis.    

Both protons and electrons are accelerated in the spherical blob but due to small radius (sub-persec) of the blob and relatively large magnetic field, the maximum energy of the accelerated cosmic rays will be limited. We assume that the particles are accelerated up to TeV energies in the blob. The shock accelerated accelerated electrons in the blazar jet are assumed to follow a broken power law energy distribution having spectral indices  $\alpha_1$ and $\alpha_2$ before and after the spectral break at Lorentz factor $\gamma_b'$ to explain the low-energy bump of the EM SED by synchrotron radiation and can be written as \cite{Banik19}
\begin{eqnarray}
N_e'(\gamma_e') = K_e \gamma_e'^{-\alpha_1} \hspace{1.5cm} \mbox{if}\hspace{0.6cm} \gamma_{e,min}' \le \gamma_e' \le \gamma_b' \nonumber \\
         = K_e \gamma_b'^{\alpha_2-\alpha_1} \gamma_e'^{-\alpha_2} \hspace{0.35cm} \mbox{if}\hspace{0.46cm} \gamma_b' <\gamma_e' \le \gamma_{e,max}'\;
\label{Eq:1}
\end{eqnarray}
where $\gamma_e' = E_e'/m_e c^2$ is the Lorentz factor of electrons of energy $E_e'$. The normalization constant $K_e$ can be found using the relation for the kinetic power of accelerated electrons ($L_e'$) in the blob frame as given in \cite{Bottcher13,Banik19}. The synchrotron radiation of the primary accelerated electrons describes the low-energy bump of the EM SED emitted from the core of Cen A and can be numerically estimated following \cite{Bottcher13, Banik19}. The inverse Compton (IC) scattering of primary accelerated electrons with the seed synchrotron photons co-moving with the AGN jet (i.e, the SSC model) describes the lower part of the high energy component (up to few GeV energies) of EM SED of the blazar Cen A following the expressions given in \cite{Banik19, Blumenthal70, Inoue96}.

Due to the large inclination angle of the jet $\theta$, such emission is not supposed to be strongly Doppler boosted. This Doppler factor is considered to be close to unity $\delta_{ob} \sim 1$ in case of Cen A. However, along the jet axis $\theta = 0^{\circ}$, the Doppler factor well be large, i.e, $\delta_j \approx 2\Gamma_j$ with any reasonable values of $\Gamma_j$. Here we consider $\Gamma_j \sim 3$ for the sub-persec jet of Cen A. Therefore, along the jet axis such emissions should be relativistically boosted which results in a much stronger emission along the jet than measured directly from Cen A. In such a case, the beamed radiation produced in the inner jet can be a dominating radiation field in kpc-scale jet region other thermal emission from an accretion disc, a broad line region (BLR) or a dusty torus (DT) for inverse-Compton scattering with relativistic electrons as discussed in next section \cite{Bednarek19}.

In the proton blazar framework, the shock accelerated cosmic rays may interact with the cold matter (protons) of density $n_{H} = (n_{e}'-{\small \displaystyle\sum_{Z,A}} Z n_A')$ in the blob of AGN jet and produce high energy gamma-rays and neutrinos. Here $n_{e}'$ and $n_A'$ are represents total number of electrons and cosmic ray nuclei (atomic number $Z$ and mass number $A$) present in that region of AGN jet respectively as described in next section. However, the contribution of hadronically generated gamma-rays can be restricted far below the inverse-Compton contribution (or Fermi$-$LAT observation) by suitable choice of cosmic ray luminosity. As for example, when cosmic ray luminosity is chosen as $3 \times 10^{42}$ ergs the gamma-rays produced in hadronic interaction is found nearly one order less than the inverse-Comption contribution when a power law energy spectrum with slope parameter $-2.5$ is considered and the acceleration efficiency is taken as $10^{-3}$. 

\subsection{The kpc-scale jet emission with proton blazar inspired model}
EM Emission from the kiloparsec-scale jet of Centaurus A has been studied from the radio, the infrared to the X-ray regime. \cite{Hardcastle06} have shown the flux observations by radio VLA (1.4, 4.9 and 8.4 GHz), infrared Spitzer (24 and 5.4 $\mu$m), ultraviolet GALEX (231 nm) and Chandra X-ray (1 keV) for three separate regions of the jet. In order to study the high-energy gamma-ray flux observed by HESS from the large scale jet (up to few kpc in projection) of the source, we consider their observations for the inner region (corresponding to 2.4$-$3.6 kpc in projection).

Similar to the core emission region, here we have considered a second moving spherical blob with a larger radius at 1 kpc away from the core of Cen-A. In our model this is the high-energy gamma-ray emission region responsible for the HESS and partially Fermi$-$LAT observations from Cen A.

The relativistic electrons and cosmic ray nuclei can be assumed to be accelerated (or re-accelerated) in the kpc-scale jet in diffusive shock acceleration by parallel collision-less shock. The accelerated relativistic electrons will obey a broken power-law energy distribution as given in Eq.~(\ref{Eq:1}) and their synchrotron emission can describe the radio to X-ray observations of the EM SED from the kpc-scale jet of the source. The synchrotron-self absorption effect is incorporated following \cite{Chen17}. The number density and energy density of relativistic (`hot') electrons are $n_{e,h}' = \int N_e'(\gamma_e') d\gamma_e'$ and $u_e'= \int m_e c^2\gamma_e' N_e'(\gamma_e') d\gamma_e'$ respectively. The synchrotron emission of the electron distribution generally describes the low-energy component of the EM SED of the AGN, which can be numerically estimated following \cite{Bottcher13, Banik19}. As elaborated in \cite{Banik19,Banik20}, the acceleration efficiency $\chi_e$ of electrons in the AGN jet is likely to be quite low and total number electrons including `hot' and non-relativistic (`cold') electrons can be determined as $n_{e}' = n_{e,h}'/\chi_e$. In the present work, we have considered the values of $\chi_e \sim 10^{-3}$, which is within the permitted range \cite{Eichler05,Banik20} and compatible with the hybrid simulation results of diffusive shock acceleration by parallel collision-less shock \cite{Gia92, Bykov96}.

The Fermi$-$LAT discovery of GeV hardness in the year 2013 and the TeV emission observed during 2004-2008 with H.E.S.S. often require a distinct, separate spectral component to explain. The inverse Compton (IC) scattering of primary accelerated electrons with the soft photon radiation field which is boosted from the core region to the in kpc-scale jet region, has been estimated  following the expressions given in \cite{Banik19, Blumenthal70, Inoue96}. The mechanism is found to contribute significantly in the observed EM SED in GeV to TeV energy range of the source. The differential population of soft photons $n_{j}'$ coming from the core region at the location of the kpc-scale jet at the distance $r$ along the jet can be estimated as \cite{Bednarek19}
\begin{eqnarray}
\epsilon_{j}'n_{j}'(\epsilon_{j}') = \frac{d_L^2\delta_j^4}{r^2 c\delta_{ob}^4} \frac{f_{\epsilon}}{m_e c^2\epsilon_{j}' }
\end{eqnarray}
where $f_{\epsilon}$ (in erg cm$^{-2}$ s$^{-1}$) represents the observed photon flux from the core of Cen A at earth, $\epsilon_{j}' = \delta_j \epsilon_{j} $ relates photon energies (in $m_ec^2$ unit) in the observer and the boosted photon energy in the kpc jet frame respectively with red shift parameter $z$, and $d_L$ is the luminosity distance between the AGN and the Earth.

In the framework of proton blazar model, the cosmic ray protons are also supposed to be accelerated in AGN jet and here we consider a mixed composition scenario for cosmic rays accelerated by Cen A. The cosmic ray energy spectrum of element atomic mass number $A$ can be assumed as $N_A'(\gamma_A') = f_A K {\gamma_A'}^{-\alpha}\exp({-\frac{\gamma_A'}{\gamma'_{A,max}}})$ where $\alpha$ is the spectral index, $\gamma_A' = \frac{E_A'}{A m_p c^2}$ is the Lorentz factor of injected cosmic rays of energy $E_A'$ in AGN jet frame and $f_A$ (for $\gamma'_{A}< \gamma'_{p,max}$) represents the relative number abundance of the accelerated cosmic ray nuclei. Also, $\gamma'_{A,min} = A \gamma'_{p,min}$ and $\gamma'_{A,max} = Z \gamma'_{p,max}$ are the minimum and maximum Lorentz factor of accelerated cosmic rays with atomic number $Z$ and mass number $A$ respectively. The normalization constant $K$ is related with the kinetic power in relativistic cosmic rays in the AGN jet frame $L_{CR}'$ via the relation \cite{Banik19}
\begin{equation}
L_{CR}' = \pi R_b'^2 \beta_j c \displaystyle\sum_{Z,A}\int_{\gamma_{A,min}'}^{\gamma'_{A,max}}m_p c^2\gamma_A' N_A'(\gamma_A') d\gamma_A'
\label{Lcr}
\end{equation}
where $n_A' = \int N_A'(\gamma_A') d\gamma_A'$ denotes the corresponding number density of individual cosmic ray nuclei and $u_{CR}' = {\small \displaystyle\sum_{Z,A}} \int m_p c^2\gamma_A' N_A'(\gamma_A') d\gamma_A'$ represents the corresponding energy density for (mixed compositions) relativistic cosmic rays in the blob of a blazar jet. 

The gamma-ray emissivity with photon energies $E_{\gamma}' = m_e c^2 \epsilon_{\gamma}'$ in the co-moving jet frame produced in cosmic ray interactions with cold proton present in the system under charge neutrality condition, $Q'_{\gamma}(\epsilon'_{\gamma}) = {\small \displaystyle\sum_{Z,A}}Q'_{\gamma,Ap} (\epsilon'_{\gamma})$ can be obtained following Eq.~(9) of \cite{Banik19} after incorporating the effect of heavier cosmic ray nuclei as given in \cite{Banik17,Banik17b,Banik18}. Due to internal photon-photon ($\gamma\gamma$) interactions as given in \cite{Aharonian08, Banik19}, the produced TeV$-$PeV gamma-rays can be absorbed while propagating through an isotropic low-frequency radiation field. The electrons/ positrons produced in $\gamma\gamma$ pair production and those produced directly due to the decay of $\pi^{\pm}$ mesons created in pp interaction will initiate EM cascades in the AGN blob via the IC scattering and the synchrotron radiation \cite{Bottcher13}. We found that this cascade emissions does not have any significant contribution to the overall spectrum in case of Cen A.   

Following \cite{Banik19} we estimated $Q_{\gamma,esc}'(\epsilon_{\gamma}')$ which denotes the total gamma-ray emissivity from the blob of AGN jet including all processes stated above i.e, the synchrotron and the IC radiation of accelerated electrons, the gamma-rays produced in $pp$ ($pA$) interaction and also the synchrotron photons of EM cascade electrons. Therefore, the observable differential flux of gamma-rays reaching at the Earth from a blazar can be evaluated from the relation
\begin{eqnarray}
E_{\gamma}^2\frac{d\Phi_{\gamma}}{dE_{\gamma}} = \frac{V'\delta_{ob}^4}{4\pi d_L^{2}}\frac{E_{\gamma}'^2}{m_e c^2} Q_{\gamma,esc}'(\epsilon_{\gamma}') . e^{-\tau_{\gamma\gamma}^{EBL}}
\label{eqgamflux}
\end{eqnarray}
where the volume of the emission region is $V' = \frac{4}{3}\pi R_b'^3$ and $E_{\gamma} = \delta_{ob} E_{\gamma}'/(1+z)$ relates the photon energies in comoving jet of redshift parameter $z$ and the observer frame, respectively. Here we have included the effect of the absorption by the extragalactic background (EBL) light on gamma-ray photons and $\tau_{\gamma\gamma}^{EBL}(\epsilon_{\gamma},z)$ is the corresponding optical depth which can be evaluated using the Franceschini-Rodighiero-Vaccari (FRV) model \cite{Franceschini08} \footnote{http://www.astro.unipd.it/background/}.

The correlated muon neutrino flux reaching at the Earth can be evaluated as 
\begin{eqnarray}
E_{\nu}^2\frac{d\Phi_{\nu_{\mu}}}{dE_{\nu}} = \xi.\frac{V'\delta_{ob}^4}{4\pi d_L^{2}} \frac{E_{\nu}'^2}{m_e c^2}Q_{\nu}'(\epsilon_{\nu}') 
\end{eqnarray}
where $Q_{\nu}'$ denotes the corresponding neutrino emissivity from the AGN jet \cite{Banik19} with energies $E_{\nu}' = m_e c^2 \epsilon_{\nu}'$ in co-moving jet frame, $\xi = 1/3$ is a fraction which is included due to neutrino oscillation and $E_{\nu} = \delta_{ob} E_{\nu}'/(1+z) $ \cite{Atoyan03} provides the relation between the neutrino energies in the observer and co-moving jet frame respectively.

The UHECRs may remain confined over a moderate time period within the AGN jet but thereafter they will slowly escape. Several instabilities grow in the process of confinement of charged particles by magnetic fields and the energetic cosmic rays will eventually, slowly escape along the field lines of the irregular field. The emitted cosmic ray particles may scatter by the extragalactic magnetic field while propagating from the source to us. The strength of the extragalactic magnetic field near the Milky Way is not clearly known. The observation of UHECR events from a sky location within $18^{\circ}$ around Cen-A indicates that the scattering is small (at least for cosmic rays above 55 EeV) and thus the magnetic field is weak. Such weak magnetic fields will only slightly enhance the travel path for Cen-A and one thus may approximate that the cosmic rays of such high energies travel straight from the source to the Earth \cite{Gopal10}. The emitted UHECRs while propagating from the source to us can suffer different interaction processes such as pair production, pion production, and photodisintegration \cite{Puget76, Allard12}. The pair production causes a continuous energy loss which starts to contribute significantly just above its energy threshold ($\sim 10^{18} eV$) while pion production starts to dominate around 70 EeV. The photodisintegration is the dominant energy loss process for UHE cosmic ray nuclei. The main background photon fields in propagation of UHECRs are cosmic microwave background and the radiation field of higher energies viz. the IR/Opt/UV background. The pair and pion production in interaction with the IR/Opt/UV is almost insignificant compare to those produced in interaction with CMB. In the nuclear photodisintegration process the nucleus is supposed to form a compound state due to photoabsorption, followed by a statistical decay process involving the emission of one or more nucleons from the nucleus. In the concerned energy range ($55 - 85$ EeV) of UHECRs the contribution of photodisintegration by CMB photons dominates over that by IR/Opt/UV backgrounds \cite{Allard12}. The expected UHECR flux reaching at the earth from nearby extragalactic sources within 75 Mpc like Cen A can be estimated as 

\begin{eqnarray}
E_{A}^2\frac{d\Phi_{A}}{dE_{A}} = \frac{V'\delta_{ob}^4}{4\pi d_L^{2}} \frac{E_{A}'^2}{m_p c^2}Q_{A}'(\gamma_{A}')e^{-\tau_{A}}
\end{eqnarray}
where $Q_{A}'(\gamma_A') =  \frac{c}{R_b'}N_A'(\gamma_A')$ represents the effective emissivity of UHECRs in AGN jet frame and $E_{A} = \delta_{ob} E_{A}'/(1+z)$ relates the cosmic ray energies in comoving jet and the observer frame, respectively. Here we have considered the possible absorption of UHECRs due to photo-meson and photo-disintegration interactions with cosmic microwave background while traveling towards us from the AGN Cen A and $\tau_{A}$ is the corresponding optical depth which can be evaluated following (\cite{Banik17b} and references therein).

The corresponding expected UHECR events within the observed energy range between $E_1$ and $E_2$ in PAO from the Cen A can be evaluated as 
\begin{eqnarray}
N_{CR} = \frac{\Xi \omega(\delta_s)}{S_{60}}{ \displaystyle\sum_{Z,A}}\int_{E_1}^{E_2}\frac{d\Phi_{A}}{dE_{A}}dE_{A}
\end{eqnarray}
where $\Xi = 20370$ km$^2$sr yr is the exposure of the detector \cite{Abreu10}, $\omega(\delta_s) = 0.64$ is the relative exposure for the declination angle ($\delta_s = -47$) and $S_{60} = \pi$ sr is the acceptance solid angle used in PAO corresponding to zenith angle $\le 60^{\circ}$ \cite{Cuoco08}.

\begin{figure*}[t]
  \begin{center}
  \includegraphics[width = 1.0\textwidth,height = 0.45\textwidth,angle=0]{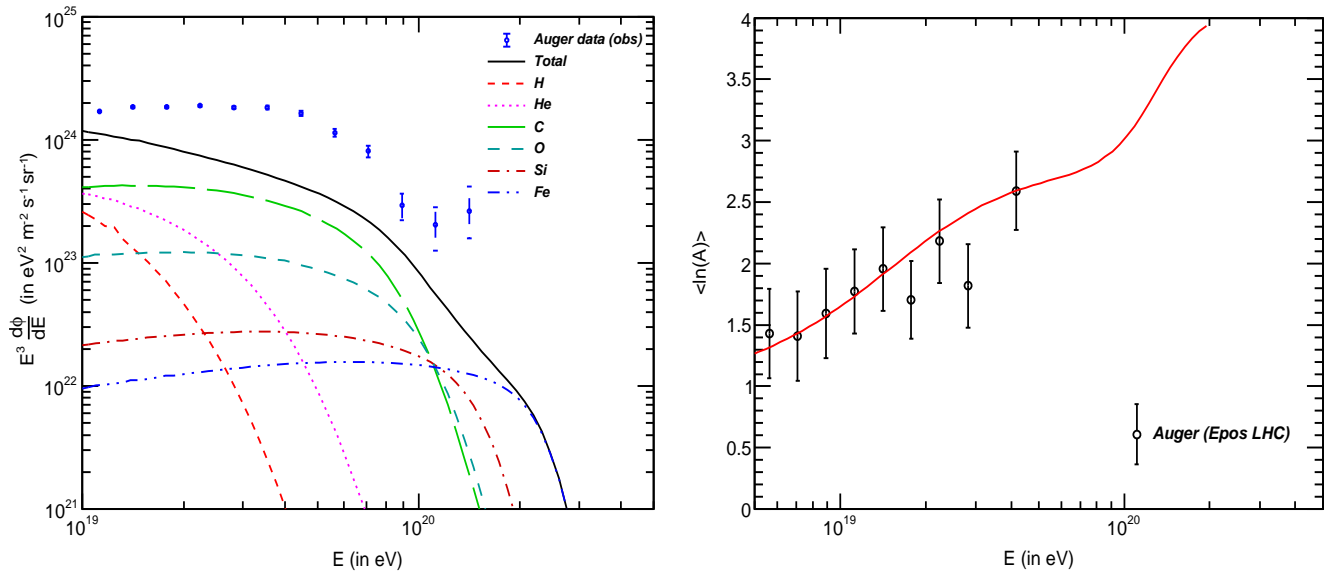}
\end{center}
  \caption{Estimated differential energy spectrum of cosmic ray neucli (left) and average logarithmic mass of mixed composition cosmic rays (right) reaching at the Earth from the radio galaxy Cen A along with Auger observation. }
\label{Fig:1}
\end{figure*}

\section{Numerical Results and discussion}
Being the nearest radio-loud active galaxy to Earth, Cen A or NGC 5128 is one of the best-studied extragalactic objects over a wide range of frequencies of its electromagnetic spectrum \cite{Israel98} and the redshift of the AGN was estimated to be $z = 0.00183$ \cite{Abdo10a}. It is the closest AGN located at a distance of 3.7 Mpc from us \cite{Ferrarese07}. The X-ray observations detected an inner kpc jet of the AGN, showing a complex structure of bright knots and diffuse radiation \cite{Kraft02}. The elliptical host NGC 5128 believed to be the remnant of a merger and features a dark lane, a thin edge-on disk of dust and young stars \cite{Aharonian09}. There is large uncertainty in the estimated values of the mass of the central black hole of Cen A, spread over almost an order of magnitude. Near-infrared stellar kinematical studies on seeing limited long-slit spectra (GNIRS at Gemini South) leads to $2 \times 10^8$ $M_{\odot}$ \cite{Silge05} which corresponds to Eddington luminosity is about $3 \times 10^{46}$ erg/s. The observed associations of 13 UHECRs as reported in 2010 \cite{Abreu10} and total 19 UHECR events \cite{Giaccari17} above 55 EeV energies with Cen A suggest that it may be one of the most thought nearby extragalactic perpetual source of cosmic rays.

As mentioned earlier, we have considered two emission region scenarios of Cen A jet and the flux estimations for the two stated scenarios are discussed below.

\subsection{The core emission}
To interpret the EM SED of Cen A over the optical to gamma-ray energy range, we have assumed the size of the blob region within the jet to be $R_b' = 2.2 \times 10^{15}$ cm with Doppler boosting factor $\delta_{ob} = 1.1$ and bulk Lorentz factor of AGN jet $\Gamma_j = 3.05$. The size of the emitting region is consistent with the size inferred from the variability, namely $R_b' \lesssim \delta_{ob} c t_{ver}/(1+z) \simeq 3.3\times 10^{15} (\delta_{ob}/1.1)(t_{ver}/10^5 s)$ cm assuming a typical variability timescale of $t_{ver}\simeq 10^5$ s. Adopted choices of Doppler boosting factor and bulk Lorentz factor of the AGN jet in core region is consistent with the upper bounds constrain of those ($\delta_{ob} \sim 1$ and $\Gamma_j \lesssim 3-5$) as given in \cite{Chiaberge01}.

The synchrotron emission of primary accelerated electron's distribution obeying a broken power law with spectral indices $\alpha_1 = 1.78$ and $\alpha_2 = 4.5$ respectively before and after the spectral break at the Lorentz factor $\gamma_b' = 1.2\times 10^{3}$ is found to explain well the lower energy bump of the experimental EM SED data of Cen A. The magnetic field of $B' = 5.6$ G and the kinetic power of primary relativistic electrons of $L_e' = 10^{42}$ erg/s in blazar jet are adopted to establish the observed low energy bump of the EM spectrum. The magnetic field used here is larger than the typical values in the core of BL Lacs which rarely exceed 1 G \cite{Tavecchio10, Inoue16}. However, several recent analysis \cite{Abdo10a, Tanada19} indicate for a very large magnetic field of a few (4$-$6) Gauss in the core region of Cen-A which is probably because of the close proximity of the core emission region of Cen-A to its nucleus \cite{Tanada19}. The primary relativistic electrons may interact with the synchrotron photons co-moving with the AGN jet via IC scattering and consequently found to produce the high-energy bump of the EM spectrum up to few GeV energies. The model fitting parameters to match the observed EM SED from the core region of the source are summarized in Table~\ref{table1}.

\subsection{The Kiloparsec-scale jet emission}
In this case, we have chosen the size of the emission region within the jet to be $R_b' = 1.9\times 10^{20}$ cm in the region of kpc-scale jet from the nucleus of the source moving with Doppler boosting factor $\delta_{ob} = 1.1$ and bulk Lorentz factor of AGN jet $\Gamma_j = 1.009$ to describe the high energy EM SED observed apparently from kpc-scale region of Cen A as well the radio to x-ray observations \cite{Hardcastle06} from the same part of the jet.  Our estimated value of bulk Lorentz factor for kpc-scale jet is consistent with the estimation from the proper motional speeds of the knots monitored by the VLA in kpc-scale jet \cite{Goodger10,Tanada19}. 

The spectral indices $\alpha_1 $ and $\alpha_2 $ of the broken power-law electron distribution are taken $2.31$ and $3.85$, respectively, before and after the spectral break at the Lorentz factor $\gamma_b' = 6.4\times 10^{5}$, which is well reproduce the lower energy bump of the experimental EM SED data of Cen A via synchrotron emission. The magnetic field and kinetic power of relativistic electrons in the blazar jet required to explain the EM spectrum are $B' = 100\times 10^{-6}$ G and $L_e' = 1.83\times 10^{42}$ erg/s, respectively, in blazar jet. The IC scattering of this primary accelerated electrons distribution with soft radiation field which is boosted from the core region as discussed above is found to contribute significantly to the measured EM spectrum by Fermi$-$LAT and HESS in Gev$-$TeV energy range. Unlike the \cite{Bednarek19}, the mechanism cannot alone describe observed GeV to TeV energy gamma-rays for this work, as we consider the appropriate accelerated electron distribution required to establish the the radio to x-ray observations from the kpc-scale jet.   

Theoretically, the maximum energy achievable in a relativistic source with the comoving size $R_b'$ and magnetic field strength, $B'$, can be obtained as $E_{A,max}' \le  ZeB'R_b' \approx Z\times 5.7\times 10^{18}$ eV in comoving jet frame \cite{Hillas84}. Here, we consider the mixed composition of primary cosmic ray nuclei with maximum energy $E_{A,max}' = Z\times 4.4\times 10^{18}$ eV that consistent with the Hillas criterion as stated above and then we proceed in the following ways:

i) Firstly, we have considered a mixed composition of cosmic rays composed of H$^1$, He$^4$, C$^{12}$, O$^{16}$, Si$^{28}$ and Fe$^{56}$ elements and we have chosed abundances ($f_A$) of various elements and their production spectrum in the jet frame in such a way to match the average atomic mass of UHECR events observed by PAO. We have estimated the average atomic mass of UHECRs for the mixed composition through the expression
$$\left<\ln(A) \right> = \frac{{ \displaystyle\sum_{z,A}}\ln(A)\frac{d\Phi_{A}}{dE_{A}}}{{ \displaystyle\sum_{z,A}}\frac{d\Phi_{A}}{dE_{A}}}$$

ii) We have taken the same value of $f_A$ and cosmic ray production spectral slope $\alpha$ ($-4.3$) \cite{Liu12} of UHECRs above 55 EeV as observed by PAO which further restrict the model parameters. 

iii) We have chosen injected cosmic ray luminosity $L_{CR}'$ in the AGN jet so as to produce observed $9.8$ UHECR events in Auger detector within the energy range between $E_1 > 55$ EeV and $E_2 = 85$ EeV from Cen A.

iv) Finally, we have computed EM (gamma-ray) spectrum given by the proton blazar inspired model using the parameters most of which are already fixed from the above considerations. 

The expected UHECR flux of different elements with their relative abundance reaching the Earth from the jet of Cen A along with their probable average atomic mass are shown in Fig.~\ref{Fig:1}. Our findings suggest a relative number abundance of the accelerated cosmic ray nuclei to be $f_H:f_{He}:f_C:f_O:f_{Si}:f_{Fe} \approx 0.54 : 0.27 : 0.15 : 3.75\times 10^{-2}:6.43\times 10^{-3}:2.68\times 10^{-3}$ (for 56 GeV $<\gamma'_{A}< \gamma'_{p,max}$) and estimated average logarithmic mass of mixed composition is found to fit consistently with the Auger collaboration estimated average logarithmic mass using Epos LHC interaction models \cite{Aab14}. The required primary cosmic ray injection luminosity is found out to be $L_{CR}' = 3\times 10^{45}$ erg/s along with the best fitted spectral slope of $\alpha = - 2.5$ which is finally cross-checked by fitting the observed high energy gamma-rays for Cen A simultaneously as discussed below. 

In the proton blazar inspired model framework, the EM spectral hardening above few GeV energies observed by Fermi$-$LAT and also HESS data of the source are found to produce well in the interactions of relativistic cosmic rays with the ambient cold protons in the blob as estimated following the equation~(\ref{eqgamflux}). Under the assumption of a low acceleration efficiency of electrons in AGN jet of $\chi_e \approx 10^{-3}$ which is within the permitted range \cite{Eichler05, Banik20} and compatible with the hybrid simulation results of diffusive shock acceleration by parallel collisionless shock \cite{Gia92, Bykov96}. Therefore, the cold proton number density in jet found out to be $1.18$ particles cm$^{-3}$ under charge neutrality condition. As discussed in the previous section and also in \cite{Banik19, Banik20}, we found that the contribution to the total EM spectrum of the IC and synchrotron emission of the stationary electron/positron pairs created in EM cascade induced by protons are very small and hence neglected.

\begin{figure}[h]
  \begin{center}
  \includegraphics[width = 0.5\textwidth,height = 0.45\textwidth,angle=0]{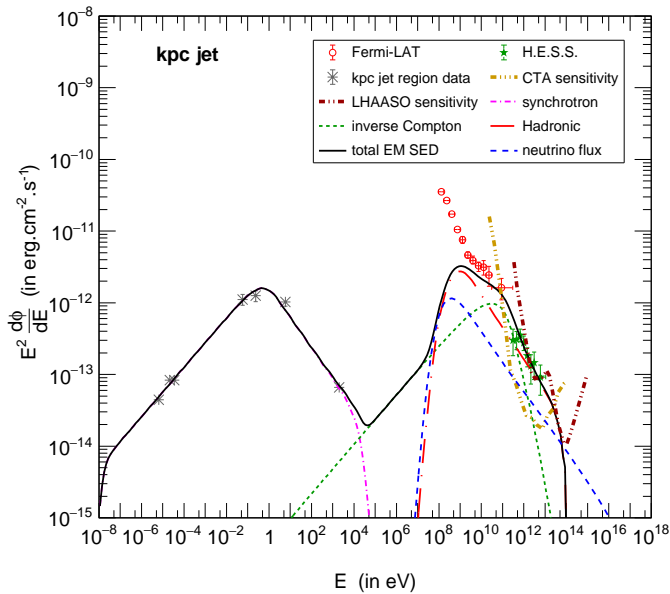}
\end{center}
  \caption{The estimated differential SED of gamma-rays and neutrinos reaching the Earth from the kpc-scale jet of radio galaxy Cen A for mixed cosmic ray composition scenario. The pink dash-single-dotted line, green dashed line denotes the EM spectrum due to the synchrotron emission of relativistic electrons and Inverse Compton scattering of relativistic electrons with inner jet photons respectively. The red large dash-single-dotted line represents the gamma-ray flux produced due to $pp$-interaction. The black continuous line represents the estimated overall differential multi wave-length EM SED. The blue long dash line indicates the differential muon neutrino flux reaching at earth. The yellow dash-triple-dotted and brown long-dash-single-dotted line denotes the detection sensitivity of the CTA detector for 1000 h and the LHAASO detector for 1 yr, respectively}
\label{Fig:2}
\end{figure}

\begin{figure}[h]
  \begin{center}
  \includegraphics[width = 0.5\textwidth,height = 0.45\textwidth,angle=0]{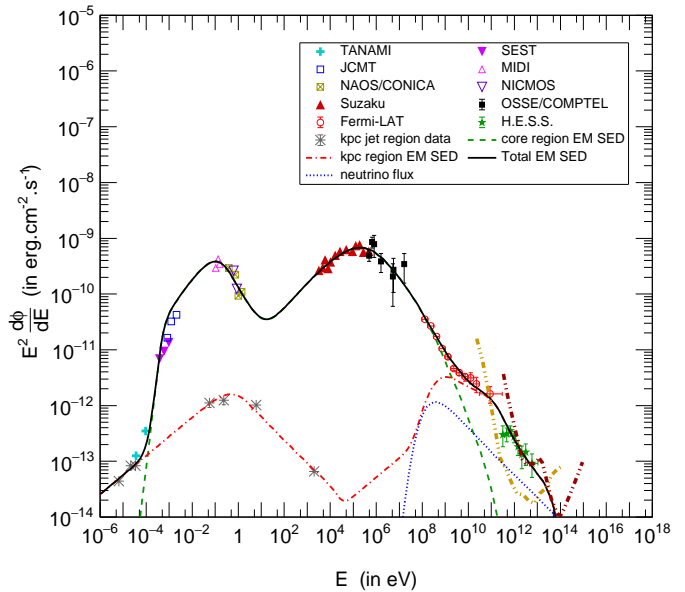}
\end{center}
  \caption{The estimated overall differential SED of gamma-rays and neutrinos from the cental part of Cen A along with different observations. The green dashed line and red dash-single dotted linere presents the gamma-ray flux produced in core region and kpc-jet region of the AGN respectively. The black continuous line denotes the estimated overall differential multi wave-length EM SED. The blue dotted line indicates the differential muon neutrino flux reaching at earth.}
\label{Fig:3}
\end{figure}

The total jet power in the form of the magnetic field, relativistic electron, cosmic rays and cold matter of density $\rho'$ (including cold protons and electrons) is evaluated via the relation \cite{Banik19}
\begin{eqnarray}
L_{jet} = \Gamma_j^2 \beta_j c \pi R_b'^2\left[\rho'c^2(\Gamma_j-1)/\Gamma_j + u' +p'\right].
\label{jetpower}
\end{eqnarray}
where $u'$ (sum of $u_e'$, $u_{CR}'$ and $u_B'$) represents comoving jet frame energy density and $p' = 1/3 u'$ (sum of $p_e'$, $p_{CR}'$ and $p_{B}'$) is the total pressure due to relativistic electrons, cosmic rays and magnetic field. The total jet power in the form of magnetic field, relativistic electron, cosmic rays and cold matter in kpc-scale jet is estimated out to be $1.14\times 10^{46}$ erg/s which is consistent with the the Eddington limit. However, there is large uncertainty in the estimated values of the mass of the central black hole of Cen A, spread over almost an order of magnitude. In Fig.~\ref{Fig:2}, we have displayed the estimated differential gamma-ray and neutrino spectrum reaching at the  Earth from this FR-I along with the different space and ground based observations. The model fitting parameters to match the EM SED as well as UHECR event are summarized in Table~\ref{table1}. The corresponding expected muon-neutrino events in the IceCube detector from the Cen A in the 32 TeV and 7.5 PeV energy range are found to be about $N_{\nu_{\mu}} = 6.4\times 10^{-2}$ events in 10 years of observations. The overall EM SED from the central part of Cen A (including core and kpc-scale jet) has been estimated and displayed in Fig.~\ref{Fig:3} along with the different experimental observations. We note that our findings can explain not only the observed multi-wavelength EM SED from the central part of Cen A but also the observed association of UHECRs with Cen A as detected by PAO.

\begin{table}[h]
  \begin{center}
    \caption{Model fitting parameters for Centaurus A.}
    \label{table1}
{\setlength{\tabcolsep}{1.em}
\renewcommand{\arraystretch}{1.2} 
    \begin{tabular}{ccc}
        \hline
      Parameters &  core region   &   kpc-scale jet   \\ \hline   \hline
      $\delta_{ob}$              &  $1.1$ &    $1.1$   \\
      $\Gamma_j$              &   $3$  &   $1.009$    \\
      $\theta$              &   $42^{\circ}$ &    $42^{\circ}$ \\
       $z$                  &  $ 0.00183$ &   $ 0.00183$\\
      $R_b'$  (in cm)            &  $2.2\times 10^{15}$ &   $1.9\times 10^{20}$ \\
       $B'$ (in G)          &  $5.6$  & $100\times 10^{-6}$  \\
       $\alpha_1$           &   $- 1.7$ &  $- 2.31$  \\ 
       $\alpha_2$           &   $- 4$ &   $- 3.85$  \\
       $\gamma_b'$  &  $1.2\times 10^{3}$ & $6.4\times 10^{5}$ \\
       $\gamma_{e,min}'$  &  $1$ &   $1$   \\
       $\gamma_{e,max}'$  &  $10^{6}$ & $10^{8}$     \\
       $L_e'$ (in erg/s)    & $10^{42}$ & $1.83\times 10^{42}$\\
       $n_H$  (in cm$^{-3}$) & $-$  & $1.18$ \\
       $\alpha_p$           &  $-$  &  $- 2.5$   \\
       $E_{p,max}'$ (in eV) & $-$  &$ 4.4\times 10^{18}$   \\ 
       $L_{CR}'$ (in erg/s)    & $-$  &$3\times 10^{45}$    \\
       $L_{jet}$ (in erg/s)   &  $1.44\times 10^{43}$  &$1.14\times 10^{46}$ \\ \hline    \hline
    \end{tabular}}\quad
  \end{center}
\end{table}

\section{Discussion and Conclusion}
The homogeneous leptonic single-zone SSC jet models usually present plausible explanations of the overall EM SED in the case of high-frequency-peaked BL Lac objects. However, the conventional one-zone SSC models can not describe satisfactorily the observed EM SED from the core of the FR-I Cen A in GeV-TeV energies \cite{Chiaberge01, Lenain08, Abdo10a}. It appears that an additional mechanism might contribute to the observed radiation at such energies \cite{Lenain08, Rieger09}. Non-simultaneous Fermi-LAT results also strongly support the said additional spectral component concept as the analysis on the core SED of Cen A shows a spectral break with photon spectral index changing from $\Gamma_{\gamma} \approx 2.7$ to $\approx 2.31$ at the break energy of $\approx 2.8$ GeV \cite{Abdalla18}. The coincident detection of the UHECRs by the PAO from Cen A provide strong support for the acceleration of cosmic rays in the AGN jet in the diffusive shock-acceleration process. The $p\gamma$ interaction and photo-disintegration models can explain the EM SED of Cen A only at TeV energies, not the overall spectrum that includes the spectral hardening at GeV energies. On the other hand, the cloud-jet interaction scenario based on $pp$ interactions is not a likely scenario for VHE emissions as no significant gamma ray flux variability has been observed. So far no consistent explanation of the EM SED from the core as well as the kpc-scale jet of Cen-A together with the observed excess of UHCR events from Cen-A and their (UHCRs) energy spectrum is available in the literature. 

In the framework of the proton blazar model, our findings imply that both the low and high-energy bump of the multiwavelength EM SED up to highest energies and also the observed UHECR events from the FR-I Cen A can be interpreted consistently considering a mixed primary composition of cosmic rays. We have not considered pure proton primary because for a consistent explanation of the observed gamma-ray and UHECRs signals from Cen A the maximum achievable energy of the source accelerated cosmic-ray protons in the observer frame is required to be the maximum observed energy of cosmic rays from Cen A, i.e. nearly 85 EeV. Though Hillas criterion allows such a maximum energy value for kpc scale jet but the PAO observations favor for heavier primaries at highest energies. For mixed composition scenario the maximum energy required in our model is (nearly) equal to the ankle energy for protons and higher energies for heavier nuclei depending on rigidities. For mixed primary composition scenario, a single power-law energy spectrum with exponential cut off with spectral index $-2.5$ needs to be adopted.

Being very sensitive up to 100 TeV energies, the coming gamma-ray experiments like CTA \cite{Ong17} and LHAASO \cite{Liu17}, may give useful evidence of the physical origin of gamma rays and the maximum achievable energy of cosmic rays in Cen A. Our estimated neutrino fluxes from the core of Cen A not only explains the non-detection of any neutrino event by Icecube observatory from Cen A but also suggests that even near-future generation of neutrino telescopes are unlikely to detect any neutrino events after a decade of observation.

\section*{Acknowledgements}
The authors would like to thank an anonymous reviewer for insightful comments and useful suggestions that helped us to improve the manuscript. AB acknowledges the financial support from SERB (DST), Govt. of India vide approval number CRG/2019/004944.

\section*{Data Availability}
In the present work, we have used the PAO measured data on the differential energy spectrum of cosmic-rays which is available on their website (also given in \cite{Verzi19}) and we have exploited the energy-dependent average logarithmic mass of observed cosmic rays as extracted by PAO considering EPOS LHC interaction models \cite{Aab14}. We have also used the data of measured electromagnetic  spectrum of Centaurus A  in the radio, optical, x-rays, and gamma-ray energies by the various observatories as given in \cite{Abdalla18}, \cite{Hess20}.

\end{document}